\documentclass[11pt,a4paper]{article}
\pdfoutput=1

\usepackage{bm}
\usepackage{amsmath}

\usepackage{jheppub}
\usepackage{amsfonts}
\usepackage{amsmath}
\usepackage{amssymb}
\usepackage{graphicx}%
\providecommand{\U}[1]{\protect\rule{.1in}{.1in}}

\newcommand{\be}{\begin{equation}}
\newcommand{\ee}{\end{equation}}
\newcommand{\bea}{\begin{eqnarray}}
\newcommand{\eea}{\end{eqnarray}}

\title{Black holes in quasi-topological gravity and conformal couplings}
\author[a]{Mariano Chernicoff,}
\author[b]{Octavio Fierro,}
\author[c,d]{Gaston Giribet}
\author[e]{and Julio Oliva.}

\affiliation[a]{Departamento de F\'{\i}sica, Facultad de Ciencias,
Universidad Nacional Aut\'onoma de M\'exico; A.P. 70-542, M\'exico D.F. 04510, M\'exico}
\affiliation[b]{Departamento de Matem\'atica y F\'{\i}sica Aplicadas,
Universidad Cat\'olica de la Sant\'{\i}sima Concepci\'on,
Alonso de Rivera 2850, Concepci\'on, Chile}
\affiliation[c]{Martin Fisher School of Physics, Brandeis University, {\it Waltham, Massachusetts 02453, USA.}}
\affiliation[d]{Departamento de F\'{\i}sica, Universidad de Buenos Aires FCEN-UBA and IFIBA-CONICET, {\it Ciudad Universitaria, Pabell\'{o}n I, 1428, Buenos Aires, Argentina.}}
\affiliation[e]{Departamento de F\'{i}sica, Universidad de Concepci\'on; Casilla 160-C, Concepci\'on, Chile.}

\date{\today}

\abstract{Lovelock theory of gravity provides a tractable model to investigate the effects of higher-curvature terms in the context of AdS/CFT. Yielding second order, ghost-free field equations, this theory represents a minimal setup in which higher-order gravitational couplings in asymptotically Anti-de Sitter (AdS) spaces, including black holes, can be solved analytically. This however has an obvious limitation as in dimensions lower than seven, the contribution from cubic or higher curvature terms is merely topological. Therefore, in order to go beyond quadratic order and study higher terms in AdS$_5$ analytically, one is compelled to look for other toy models. One such model is the so-called quasi-topological gravity, which, despite being a higher-derivative theory, provides a tractable setup with $R^3$ and $R^4$ terms. In this paper, we investigate AdS$_5$ black holes in quasi-topological gravity. We consider the theory conformally coupled to matter and in presence of Abelian gauge fields. We show that charged black holes in AdS$_5$ which, in addition, exhibit a backreaction of the matter fields on the geometry can be found explicitly in this theory. These solutions generalize the black hole solution of quasi-topological gravity and exist in a region of the parameter spaces consistent with the constraints coming from causality and other consistency conditions. They have finite conserved charges and exhibit non-trivial thermodynamical properties. }

\emailAdd{mchernicoff@ciencias.unam.mx}
\emailAdd{gaston@df.uba.ar}
\emailAdd{julioolivazapata@gmail.com}
\emailAdd{ofierro@ucsc.cl}

\begin{document}

\maketitle
\flushbottom

\section{Introduction}

In the last decade there have been a lot of interesting work on the effects produced by higher-curvature corrections in the context of the AdS/CFT correspondence. The most celebrated example is the observation \cite{Brigante, Brigante2} that $R^2$ terms in 5 spacetime dimensions allow for a violation of the Kovtun-Son-Starinets (KSS) bound \cite{KSS} in the dual CFT$_4$. In this and related contexts, holography appears as a method to constrain the higher-curvature couplings in the bulk theory by considering consistency conditions of the boundary theory, and vice versa. For instance, $R^2$ terms have been investigated in relation to the so-called conformal collider physics \cite{MH, Hofman}, where the relation between constraints imposed by causality in the bulk of AdS$_5$ and the condition of positivity of the energy in the dual CFT$_4$ were studied. This translates into bounds for the ratio of central charges $a/c$ in the CFT$_4$, which can be computed using holographic renormalization in the bulk theory \cite{HS}. Higher-curvature terms in the context of AdS$_D$/CFT$_{D-1}$ have also been studied in higher-dimensions, $D > 5$ \cite{Camanho, deBoer, Camanho2, Camanho3, Camanho4}. It turns out that, in all these examples, the favourite playground to perform the analytic computations is Lovelock theory. Being its field equations of second order, Lovelock theory permits to study several setups, including black holes analytically. This is why the $R^2$ terms studied in \cite{Brigante, Brigante2, Camanho}, as well as the cubic \cite{GeSin,deBoer, Camanho2, Camanho3} and quartic \cite{Camanho4} generalizations studied subsequently, are all particular examples of Lovelock theory. 

This game, however, has an obvious limitation: It turns out that second-order terms with $R^n$ couplings only exist in dimension $D\geq 2n+1$, while in $D< 2n+1$ dimensions the inclusion of non-trivial higher-curvature terms necessarily demands higher-derivative couplings or non-minimal couplings with additional matter fields. Therefore, if one is interested, for instance, in investigating $R^3$ terms in $D=5$ dimensions, one is compelled to look for another tractable model different from Lovelock theory. More concretely, consider the higher-curvature theories in $D$ spacetime dimensions defined by augmenting the Einstein-Hilbert action with terms of the form   
\begin{equation}
\Delta I^{(n)} = a_n \  \int d^Dx \sqrt{-g} \ \ t_{a_1 b_1 a_2 b_2 \ ... \ a_n b_n }^{c_1 d_1 c_2 d_2 \ ... \ c_n d_n } \ R_{c_1 d_1}^{\ \ \ \ a_1 b_1 }   ... \ R_{c_n d_n}^{\ \ \ \ a_n b_n }  + I_B,\label{Love}
\end{equation}
where $a_n$ is a coupling constant, $t_{a_1...b_n }^{c_1...d_n }$ is a tensor that does not depend on the fields, and $I_B$ is a boundary term. We could also consider terms of the same dimension containing derivatives of the Riemann tensor $\nabla^n R$, but we will ignore such couplings here. Properties of the theory are encoded in the tensor $t_{a_1...b_n }^{c_1...d_n }$, which has the appropriate symmetries and it is defined up to changes in $I_B$. In the particular case
\begin{equation}
t_{a_1 b_1 \ ... \ a_n b_n }^{c_1 d_1 \ ... \ c_n d_n } = \frac{(2n)!}{2^n} \ \delta^{c_1}_{[a_1}  \delta^{d_1}_{b_1} \ ... \ \delta^{a_n}_{c_n}\delta_{b_n]}^{d_n} \label{JKL}
\end{equation}
the terms in (\ref{Love}) correspond to the Lagrangian of Lovelock theory, which in $D=2n$ are total derivatives and in $D<2n$ identically vanish. Then, the question arises as whether it is possible to find in $D$ dimensions a non-trivial tractable model other than (\ref{Love}) that shares with the latter its nice properties and includes, in addition, $R^n$ terms with $n>D/2$. Of course, the answer depends on what does one mean by ``tractable model'' and ``nice properties''. Suppose, for example, that one is interested in black holes in asymptotically AdS space; then, a more concrete question is whether a tractable model with $R^n$ terms exists in dimension $D<2n$ for which one can study black holes analytically. Remarkably, such theories do exist. For $R^3$ terms in $D=5$ dimensions a theory dubbed quasi-topological gravity has been presented in \cite{OR, MR, Myers}, which is a particular case of a larger family of theories studied in \cite{OR2}. Among other interesting properties, such as exhibiting a GR like massless spin-2 propagating graviton, quasi-topological gravity admits black hole solutions with interesting thermodynamical features and whose metrics can be written explicitly. In this paper, we will study and generalize these black hole solutions by investigating to what extent the nice properties of Lovelock black holes are also exhibited by the solutions of quasi-topological gravity. For instance, it is known that, in an arbitrary number of dimensions, Lovelock gravity conformally coupled to scalar matter admits simple black hole solutions that present a backreaction of the matter on the geometry \cite{GLOR}. These solutions represent black holes with a secondary hair described by a scalar field configuration that is regular everywhere outside and on the horizon, and decays at large distance in a way compatible with the AdS asymptotic. The metric of these black holes can be written explicitly for Lovelock theory including $R^2$, $R^3$, and $R^4$ terms. Here, we will show that the same happens in quasi-topological gravity. We will find explicit solution of charged AdS black holes in quasi-topological gravity conformally coupled to a scalar field, and in presence of an Abelian gauge field, which exhibit a backreacting scalar field configuration that is regular everywhere outside and on the horizon. We will compute the conserved charges and study the thermodynamical properties of these black holes, and show they describe interesting physics. 

We organize the paper as follows: In section II, we review quasi-topological gravity. We present it in a way that is convenient for our computation. We also introduce the matter content of the theory, which consists on an Abelian gauge field and a real scalar field conformally coupled. In section III, we derive the black hole solutions, which present both electric charge and backreaction due to the presence of the scalar field. We compute their conserved charges as well as their associate thermodynamical quantities. In section IV, we generalize the solution by adding $R^4$ terms in the gravitational action. The inclusion of $R^4$ terms also yields an analytically solvable model. In section V, we generalize our result by considering more general type of coupling with the scalar matter, for which we also construct explicit black hole solutions. We conclude in section VI with some final remarks.

\section{Quasi-topological gravity}

\subsection{The gravitational action}

We will consider a theory defined by an action of the form
\begin{equation}
I\left[  g_{ab},A_{a},\chi \right]  = I_{\text{gravity}}\left[  g_{ab} \right] + I_{\text{matter}}\left[  g_{ab},A_{a},\chi \right]\ , \label{TheTheory}
\end{equation}
where the first term on the right hand side represents the action containing uniquely the metric as a field. This purely gravitational part of the action is given by the quadratic Einstein-Gauss-Bonnet action plus a very precise combination of cubic scalar invariants; namely
\begin{equation}
I_{\text{gravity}}\left[  g_{\mu\nu} \right] =\int d^{5}x \sqrt{-g} \left(  \frac{R-2\Lambda}{16\pi G}+a_{2}\mathcal{L}_{2}  +a_{3}\mathcal{L}_{3}\right)  \ , \label{La4}
\end{equation}
with the quadratic combination given by
\begin{equation}
\mathcal{L}_{2}=  R^{2}-4R_{ab}R^{ab}+R_{ab}^{\ \ cd}R_{cd}^{\ \ ab} \ ,
\end{equation}
and the cubic combination given by
\begin{multline}
\mathcal{L}_{3}=-\frac{7}{6}R_{\ \ cd}^{ab}R_{\ \ bf}^{ce}R_{\ \ ae}%
^{df}-R_{ab}^{\ \ cd}R_{cd}^{\ \ be}R_{\ e}^{a}-\frac{1}{2}R_{ab}%
^{\ \ cd}R_{\ c}^{a}R_{\ d}^{b}\\ +\frac{1}{3}R_{\ b}^{a}R_{\ c}^{b}R_{\ a}%
^{c}-\frac{1}{2}RR_{\ b}^{a}R_{\ a}^{b}+\frac{1}{12}R^{3} \label{L3OR}
\end{multline}
where $a_2$, $a_3$ are coupling constants. Hereafter we set $16\pi G=1$. This is the most general theory whose traced
field equations lead to a second order constraint on the metric and,
furthermore, with field equations that are of second order for spherically
(planar or hyperbolic) symmetric spacetimes. As introduced in \cite{OR},
$\mathcal{L}_{3}$ is obtained by setting $D=5$ and $n=3$ in the following expression%
\begin{multline}
\mathcal{L}_{n}={\frac{1}{2^{n}}}\left(  \frac{1}{D-2n+1}\right)
\delta_{c_{1}d_{1}\cdots c_{n}d_{n}}^{a_{1}b_{1}\cdots a_{n}b_{n}}\left(
C_{a_{1}b_{1}}^{c_{1}d_{1}}\cdots C_{a_{n}b_{n}}^{c_{n}d_{n}}-R_{a_{1}b_{1}%
}^{c_{1}d_{1}}\cdots R_{a_{n}b_{n}}^{c_{n}d_{n}}\right)   \\ -c_{n}C_{a_{1}b_{1}%
}^{a_{n}b_{n}}C_{a_{2}b_{2}}^{a_{1}b_{1}}\cdots C_{a_{n}b_{n}}^{a_{n-1}%
b_{n-1}} \label{Lk}%
\end{multline}
where $C_{a \  c d}^{\ b}$ is the Weyl tensor and
\begin{equation}
c_{n}={\frac{(D-4)!}{(D-2n+1)!}}{\frac
{[n(n-2)D(D-3)+n(n+1)(D-3)+(D-2n)(D-2n-1)]}{[(D-3)^{n-1}(D-2)^{n-1}%
+2^{n-1}-2(3-D)^{n-1}]}\ .}%
\end{equation}
Writing the Weyl tensor in terms of the Riemann tensor, expression (\ref{Lk}) can
be shown to be well-defined for $D=2n-1$. It can also be easily shown that 
\begin{equation}\label{L3}
\mathcal{L}_{3}=\frac{7}{24}\mathcal{Z}^{\prime}%
+\frac{7}{48}E_{6} 
\end{equation}
where $\mathcal{Z}^{\prime}$ is the quasi-topological
combination introduced in \cite{MR} in 5 dimensions, and $E_{6}$ is the 6-dimensional Euler density
\begin{eqnarray}
E_6 &=&R^{3}+3RR^{mn ab }R_{abmn
}-12RR^{mn }R_{mn}+24R^{mnab }R_{am
}R_{bn }+16R^{mn }R_{na }R_{m }^{a }+  \nonumber \\
&&+24R^{mnab }R_{abnr }R_{m }^{r}+8R_{\ \ ar }^{mn }R_{\ \ ns }^{ab }R_{\ \
mb }^{rs }+2R_{abrs }R^{mnab }R_{\ \ mn}^{rs },  \label{cubo}
\end{eqnarray}
 which identically vanishes in $D=5$. --{Our notation relates to that in \cite{MR} as follows: $\Lambda = -6/L^2$, $a_2 = \lambda L^2/2$, $a_3=3\mu L^4$, $f_{\infty }=|\lambda |L^2/6$.}-- Writting
$\mathcal{L}_{3}$ as in equation (\ref{L3OR}) is advantageous for several reasons: On the one hand, at linearized level around a conformally flat background --in
particular maximally symmetric (A)dS spacetime--, the equations are manifestly of second
order since potentially dangerous terms in the field equations are quadratic
in the Weyl tensor and therefore do not contribute to linear order around AdS. On the other hand, the equations evaluated on a spherically symmetric ansatz render of second order
since all the complete contractions of Weyl tensors are proportional to a
single combination and the relative factor between $\delta^{\left(  3\right)
}C^{3}$ and $tr\left(  C^{3}\right)  $ is such that on spherically symmetric
spacetimes that combination vanishes. Furthermore, equation (\ref{L3}) gives a clear way to generalize this structure to Lagrangian of order $n$ in the curvature
in dimensions $D=2n-1$.

There are other theories with $R^3$ terms that can be written in 5 dimensions and in which black holes can be studied; see for instance the recent works \cite{Bueno, Bueno2, Bueno3, Mann}.   

\subsection{Matter content and conformal couplings}

The matter content of the theory consists of an Abelian gauge field, which action is given by
\begin{equation}
I_{U(1)}=-\frac{1}{4}\int d^{5}x\sqrt{-g}\ F_{ab}F^{ab}\ .
\end{equation}
and a conformally coupled real scalar field $\chi$·. The latter couples non-minimally to linear and quadratic terms of the curvature. We consider the most general theory for a conformal scalar which leads to second
order field equations in 5 dimensions \cite{PutoElQueLee}. The action for such a theory takes the following form,%
\begin{equation}
I_{\chi }=\int d^{5}x\sqrt{-g}\left[  b_{0}\chi^{-5/s}+b_{1}\chi^{-\left(
3+2s\right)  /s}\hat{R}+b_{2}\chi^{-\left(  1+4s\right)  /s}\left(  \hat{R}^{2}%
-4\hat{R}_{ab}\hat{R}^{ab}+\hat{R}_{abcd}\hat{R}^{abcd}\right)  \right]  \ ,
\end{equation}
where $b_0$, $b_1$, $b_2$ are coupling constants, $s$ the conformal weight, and where
\begin{equation}
\hat{R}_{ab}^{\ \ cd}\equiv\chi^{2}R_{ab}^{\ \ cd
}+\frac{4}{s}\chi\delta_{\lbrack a}^{[ c}\nabla_{b]}\nabla
^{d]}\chi+\frac{4(1-s)}{s^{2}}\delta_{\lbrack a}^{[c}%
\nabla_{b]}\chi\nabla^{d]}\chi-\frac{2}{s^{2}}\delta_{\lbrack a
}^{[c}\delta_{b]}^{d]}\nabla_{e}\chi\nabla^{e}\chi,
\label{Stensor}%
\end{equation}
transforms covariantly under local Weyl rescalings%
\begin{equation}
g_{ab}\rightarrow e^{2\sigma\left(  x\right)  }g_{ab},\qquad
\chi\rightarrow e^{s\sigma\left(  x\right)  }\chi\,.
\label{conformaltrans}%
\end{equation}
More precisely, under the rescaling (\ref{conformaltrans}) tensor
(\ref{Stensor}) transforms as
\begin{equation}
\hat{R}_{ab}^{\ \ cd}\rightarrow e^{2(s-1)\sigma\left(  x\right)
}\hat{R}_{ab}^{\ \ cd}\,.
\end{equation}
With these ingredients one can directly prove that the matter action is
invariant under (\ref{conformaltrans}).

\section{Quasi-topological black holes}

\subsection{Hairy black holes}

Let us fix for future convenience the conformal weight of the scalar to be $s=-1/5$. Theory (\ref{TheTheory}) admits a hairy, charged black hole solution where the scalar field
is given by
\begin{equation}
\chi=\frac{n}{r^{1/5}}\ , \label{La15}
\end{equation}
the vector potential reads%
\begin{equation}
A_{a}=\frac{\sqrt{3}e}{r^{2}}\delta_{a}^{t}\ , \label{La16}
\end{equation}
and the metric takes the form
\begin{equation}
ds^{2}=-N\left(  r\right)  ^{2}f\left(  r\right)  dt^{2}+\frac{dr^{2}%
}{f\left(  r\right)  }+r^{2}g_{AB}dx^A dx^B\ .\label{anz}%
\end{equation}
Here $g_{AB}$ ($A,B=1,2,3$) is the metric of an Euclidean three-dimensional
manifold of constant curvature $k$, which can be normalized to
$k=0, \pm1$. When the coupling constants of the matter theory satisfy the relation \cite{GLOR}
\begin{equation}
b_{2}=\frac{9b_{1}^{2}}{10b_{0}}\ , \label{La19}
\end{equation}
there exists a solution with constant lapse function $N(r)$. Here, we consider $N  =1$, which corresponds to a particular choice of the value of the speed of light $c$ at the conformal boundary. The function $f\left(  r\right)
=k+g\left(  r\right)  $ in the metric gets fixed by the following Wheeler polynomial 
\begin{equation}
2a_{3}r^{2}g^{3}+12a_2 r^{4}g^{2}-6r^{6}g-\Lambda r^{8}-6(2a_2 |k|%
+\mu)r^{4}-6qr^{3}+6r^{2}e^{2}=0\ , \label{wp}%
\end{equation}
where $\mu $ and $e$ are two integration constants, while $q$ is related to the scalar field strength by
\begin{equation}
q=\frac{n^{15}}{3}(b_{0}n^{10}+6b_{1}k)\ .
\end{equation}
Constants $\mu $ and $e$ are associated to the mass and the electric charge respectively (see (\ref{LaMasa}) and (\ref{LaCarga}) below). Similarly, $q$ can be regarded as a charge associated to the scalar field. However, unlike $\mu $ and $e$, which are arbitrary, $q$ has an absolute value fixed by the couplings of the theory%
\begin{equation}\label{eln}
n=\varepsilon\left(  -\frac{18}{5}\frac{b_{1}}{b_{0}}k\right)  ^{1/10}\ .
\end{equation}
This means that this is actually a secondary hair, or more precisely a discrete charge that can take three values $q=0,\pm |q|$; namely
\begin{equation}\label{elq}
q=\frac{4kb_{1}}{5}\varepsilon\left(  -\frac{18b_{1}k}{5b_{0}}\right)
^{3/2}\ ,
\end{equation}
with $\varepsilon=0,\pm1 $.

\subsection{The AdS$_5$ vacua}

The maximally symmetric solutions of the theory are locally defined by metrics
of the form (\ref{anz}) with $f\left(  r\right)  =-\frac{\lambda}{6}r^{2}+k$,
and vanishing matter fields. The curvature of the solution is given by the dressed cosmological constant $\lambda $, which is given by the roots of the cubic polynomial
\begin{equation}
a_{3}\lambda^{3}-36a_2\lambda^{2}-108\left(  \lambda-\Lambda\right)  =0
\end{equation}
with
\begin{equation}
\mu=-2|k|a_2\ . \label{La24}
\end{equation}

Equation (\ref{La24}) expresses a mass-gap that typically appears when quadratic terms are included in
dimension 5 --similarly as it happens in GR in 3 dimensions--.

There are three possible solutions for $\lambda $, but only one of
those defines the GR branch, meaning the branch for which
$\lambda\rightarrow\Lambda$ as $a_{3}$ and $a_2$ go to zero. In fact, the first restrictions on the parameter space comes from demanding that the root $\lambda $ associated to the GR branch to be negative, in order to have a perturbative AdS vacuum. A second constraint follows from demanding the absence of ghosts around the AdS$_5$ vacuum: Defining the length scale $L =\sqrt{ -6/\Lambda } $, perturbations around AdS$_5$ yield the condition
\begin{equation}
1-\frac{2}{3} a_2 |\lambda | - \frac{1}{36} a_3 |\lambda |^2 > 0 \label{noghosts}
\end{equation}
to be ghost-free. 

Further consistency constraints come from holography, as the central charges ratio $a/c$ has to obey certain bounds \cite{MH}; see also references thereof. The holographic Weyl anomaly \cite{HS} computed for this model yields the central charges
\begin{eqnarray}
a = \frac{2\pi^2}{|\lambda / 6|^{3/2}} (1-4a_2 L^{-2} -a_3 L^{-4} ) \ , \ \ \ 
c = \frac{2\pi^2}{|\lambda / 6|^{3/2}} (1-12a_2 L^{-2} +3a_3 L^{-4} ) 
\end{eqnarray}
which enter in the expression of expectation value of the CFT$_4$ stress tensor
\begin{equation}
\langle T_{\mu }^{\ \mu }\rangle = \frac{c}{16\pi ^2} W_4 - \frac{a}{16\pi ^2} E_4   
\end{equation}
where $E_4=R_{ijkl}R^{ijkl}-4 R_{ij}R^{ij} + R^2$ is the 4-dimensional Euler density, and $W_4=R_{ijkl}R^{ijkl}-2 R_{ij}R^{ij} +(1/3) R^2$ is the Weyl invariant combination. 

Remarkably, the existence of black hole solutions (\ref{anz})-(\ref{wp}) are compatible with all the constraints of the coupling constants $a_2$, $a_3$, $\Lambda $.

\subsection{Conserved charges}

As a preliminary ingredient for the analysis of the thermodynamics, we are interested in computing the conserved charges associated to the black hole solution. The mass and
electric charged can be obtained by means of the Regge-Teitelboim approach \cite{GGO}, i.e. by
finding the variation of the boundary term of the action that renders a well
defined variational principle \cite{Nosotros}, and reading the mass and the electric charge
from the terms multiplying the lapse function $N(r)$ and the electric potential
at infinity. This yields the following value for the mass
\begin{equation}
{M}  =\frac{\text{Vol}}{2r_{+}^{2}}\left(  -\Lambda r_{+}^{6}+6r_{+}^{4}%
k-6qr_{+}+6e^{2}-2a_{3}k\right)   , \label{LaMasa}
\end{equation}
where $r_+$ is the horizon radius and $\text{Vol}$ is the volume of the unit radius constant curvature base manifold with metric $g_{ab}$. In other words, we have ${M} = 3\mu \text{Vol}$. Mass (\ref{LaMasa}) is actually defined up to a constant $M_0$, which corresponds to the gravitational energy of the reference background. Since the theory has a mass-gap and also exhibits an extremal solution, there is more than one natural candidate to fix the value of $M_0$. 

In a similar manner we can compute the electric charge, which is seen to be
\begin{equation}
{Q}=-\sqrt{3}e \text{Vol} . \label{LaCarga}
\end{equation}

\subsection{Thermodynamics}

The Euclidean approach is also useful to obtain the entropy $S$ from the
boundary term at the horizon, which coincides with the expression for the
entropy obtained from the Wald formula \cite{Wald}. This reads
\begin{equation}
\frac{S}{\text{Vol}}=4\pi {r_{+}^{3}}+48\pi ka_2 r_{+}-10\pi k q+\frac{12\pi |k|a_{3}%
}{r_{+}}\ . \label{LaSusana}
\end{equation}

The leading contribution to $S$ for sufficiently large $r_+$ is given by the Bekenstein-Hawking area term --recall that in our notation $1/(4G)=4\pi $--. There are, in addition, subdominant contributions due to the presence of higher-curvature couplings. In particular, there is a constant term, which comes from the non-minimal couplings with the scalar field. The latter contribution is somehow analogous to the entropy contributions of topological $R^n$ terms in $D=2n$ dimensions.

As usual, the Hawking temperature can be obtained as the inverse of the period of the Euclidean time required for the absence of conical singularities in the
Euclidean continuation of the geometry. This yields
\begin{equation}
T=\frac{-2\Lambda r_{+}^{6}+6kr_{+}^{4}+3qr_{+}+(2a_{3}k-6e^{2})}%
{12\pi r_{+}\left( r_{+}^{4}+4a_2 kr_{+}^{2}-a_{3}|k|\right)  } . \label{LaT}
\end{equation}%

As expected, when $a_2$ and $a_3$ are zero, this expression coincides with the temperature of GR black holes. The presence of $R^2$ and $R^3$ terms change the behaviour in the UV. The scalar hair also changes the qualitative behaviour of the solution at short distances, inducing a change in the specific heat for black holes with radius significantly smaller than the scale $\sim |q|^{1/3}$. This implies that the theory leads to remnants. In fact, among the peculiar features of (\ref{LaT}) is that it allows for black holes with arbitrarily low temperature in AdS$_5$; there exists an extremal solution with a degenerate horizon for which $T$ vanishes. The near-horizon region of this extremal black hole is $AdS_2 \times S^3$.

Considering independent variations of the continuous parameters $r_{+}$ and
$e$, one can check that the first law of black hole thermodynamics is indeed
fulfilled; namely
\begin{equation}
dM=TdS+\Phi dQ\ , \label{Elprimero}
\end{equation}
where the horizon electric potential is given by $\Phi=-{2\sqrt{3}e}/{r_{+}^{2}}$.

\subsection{The physics behind the parameter space}

Given the amount of coupling constants and parameters, the thermodynamical variables computed above may exhibit quite different qualitative behaviors. To study the different possibilities, let us begin by considering the case of spherical horizons $k=1$. Some comments are in order:

\begin{itemize}

\item Formula (\ref{LaMasa}) shows that when $a_3=0$ the mass $M$, taken as a function of $r_+$, is not bounded from below if $e=0$ and $q>0$. Furthermore, even in the case $e\neq 0$, the solution with positive $q$ exhibits other pathologies such as negative entropy configurations \cite{GGGO}. This is consistent with the fact that, for positive $q$, the energy density associated to the scalar field becomes negative. This means that it is sensible to restrict the analysis to values $q\leq 0$. In the case $a_3>0$ something similar occurs due to the presence of the last term in (\ref{LaMasa}). This suggests that one should impose the condition $a_3<3e^2$. Nevertheless, this feature is cured when higher-curvature corrections are taken into account. For instance, we will see that $R^4$ terms with positive coupling constant suffice to render the formula for $M$ positive for all values of $a_3$. 

\item In addition, the pathologies that could potentially develop at short distances are not necessarily a problem provided something drastic happens at certain critical radius. In that case, the thermal evaporation can probably stops before the smaller scales are reached. For example, if the temperature $T$ vanishes at a given radius $r_{\text{ext}}$, after having undergone an evaporation stage with positive specific heat, the black hole yields a remnant of finite radius. Extremal solutions with $T=0$ exist in this theory even for $e=0$. This is a typical feature of higher-curvature black holes; however, here this is also possible due to the presence of the backreacting scalar field. If $q<0$, temperature vanishes for black holes with radius $-2\Lambda r_{\text{ext}}^{6}+6r_{\text{ext}}^{4}+3qr_{\text{ext}}+(2a_{3}-6e^{2})=0$, which takes the minimal value for $e=0$. In this case, the solution exhibits a degerate horizon close to which the geometry becomes AdS$_2 \times S^3$.

\item Temperature may also diverge for positive values of $r_+$. Whether or not this happens at a radius larger than the extremal radius $r_{\text{ext}}=r_+^{(T=0)}$ depends on the range of parameters. In fact, if $a_3>0$ the temperature develops a pole at $r_+^{(T=\infty )}=\sqrt{\sqrt{4a_2^2+a_3}-2a_2}.$ This infinitely hot solutions of finite radius may seem puzzling; one may restrict the space of parameters in such a way that the minimal extremal radius $r_{\text{ext}}^{(e=0)}$ has to be large than $r_+^{(T=\infty )}$.

\item Formula (\ref{LaSusana}) for the entropy also exhibits curious features. In particular, it may become negative for certain range of parameters. More importantly, because of the last term in (\ref{LaSusana}), it increases inversely proportional to the radius for sufficiently small black holes. This is a phenomenon that does not occur in the case of Lovelock black holes, and can be regarded as a direct consequence of having $R^n$ terms in dimension lower than $2n+1$. We will see below that this becomes even more drastic when $R^4$ terms are included, as in that case the entropy increases inversely proportional to the cube of the horizon radius.
 
\end{itemize} 

In order to give an example of how the different signs and values of the coupling constants and charges enter in the contributions of the thermodynamical variables, let us consider the following sequence: Suppose one starts by considering negative values of $a_3$, positive $a_2$ and no charges, i.e. $e=q=0$. In this case, $M$ turns out to be bounded from below and $T$ does not exhibit poles. However, in this case the entropy takes negative values for sufficiently small radius. In order to avoid the latter feature, one can propose to reverse the sign of $a_3$ and introduce a negative value for $q$. This would make $S$ to be positive definite; however, this would also bring us back to the situation in which $T$ presents a pole at certain radius and, more importantly, $a_3 >0$ renders the mass unbounded from below. To solve the latter, one may propose to introduce $R^4$ terms with appropriate sign for the coupling constant ($a_4>0$ below). In fact, this makes $M$ bounded from below again; however, for the sign of $a_4$ for which the mass spectrum is positive definite, the entropy receives an additional negative contribution coming from the $R^4$ terms. This pattern shows that the addition of higher-curvature terms, while solving pathologies that may appear for $M$ and $T$, typically gives unphysical contributions to $S$. This is solved as suggested above; namely by considering only black hole configurations such that the horizon radius satisfies the hierarchy $r_+ \geq r_+^{(T=0)} \geq r_+^{(S=0)}$. This is always possible in a way that it is consistent with $\lambda <0$; see the figure below. The figure shows two qualitatively different scenarios in which this hierarchy is respected. The reason why different behaviors are possible is that there are several length scales involved in the problem, which are given by the dimensionful coupling constants. These scales compete and give rise to different physics depending on the region of the parameter space. The figure shows the typical behavior of the temperature and entropy of a higher-curvature black hole as functions of its horizon radius. At large distances, these quantities behave as those of AdS black holes in $D=5$ GR, namely $T\sim r_+ $, $S\sim r_+^3 $. At short distances, in contrast, the behavior is qualitatively different from GR. In particular, for the choice of parameters corresponding to the graph on the left, the temperature takes a maximum value at a critical value below which the black hole exhibits negative specific heat. The temperature eventually vanishes for a given extremal radius $r_+^{(T=0)}$ due to the presence of both the higher-curvature terms and the scalar field backreaction. For suitable choice of the parameters, this extremal radius results larger than the radius $r_+^{(S=0)}$ below which the entropy becomes negative. Notice also that whether or not there exist black hole solutions with positive specific heat depends on the values of the coupling constants. For instance, the graph on the right shows a case in which the scale $|a_3|^{1/4}$ dominates over the scale $|q|^{1/3}$ and settles down between the extremal radius $r_+^{(T=0)}$ and the cosmological scale $|\lambda |^{-1/2}$. This strangulation scale excludes the possibility of solutions with negative specific heat and all the AdS black holes turn out to be ``large''. This is similar to what happens in Chern-Simons gravity --including $D=3$ GR--.

In the case of horizons with negative constant curvature, $k=-1$, whose geometrical interpretation is more elusive, several of the features discussed above are also present. In fact, both the boundedness of $M$ and the vanishing of $T$ actually depend on $k$ and on the product $a_3k$, but not on $a_3$ separately. In contrast, the pole condition of $T$ does not depend on the sign of the horizon curvature and of the coupling constant $a_3$ independently. In the case of vanishing horizon curvature, $k=0$, the scalar field vanishes and the analysis reduces to that of \cite{MR}. 

\begin{center}
\begin{figure}
\includegraphics[scale=0.28]{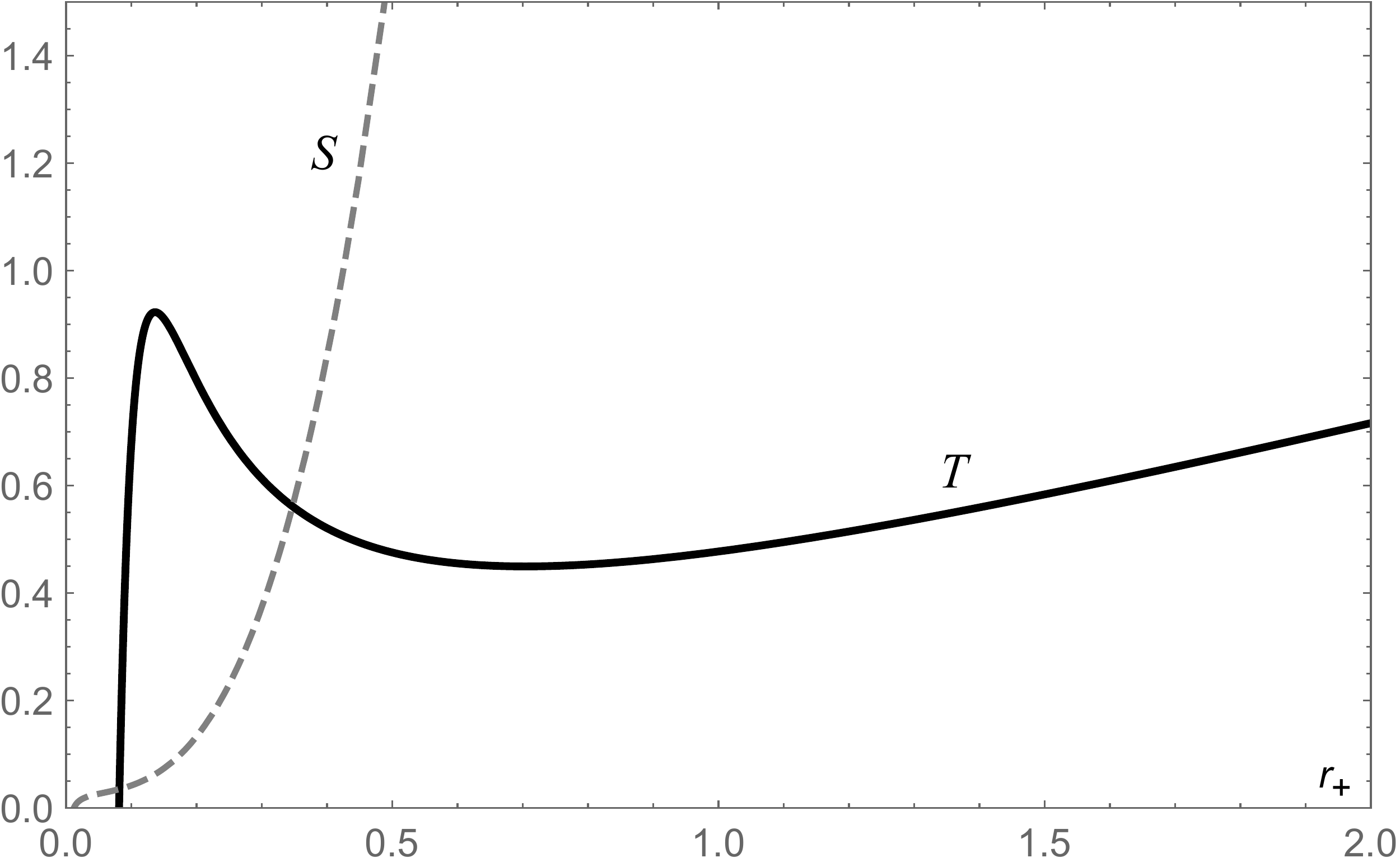}\quad\includegraphics[scale=0.28]{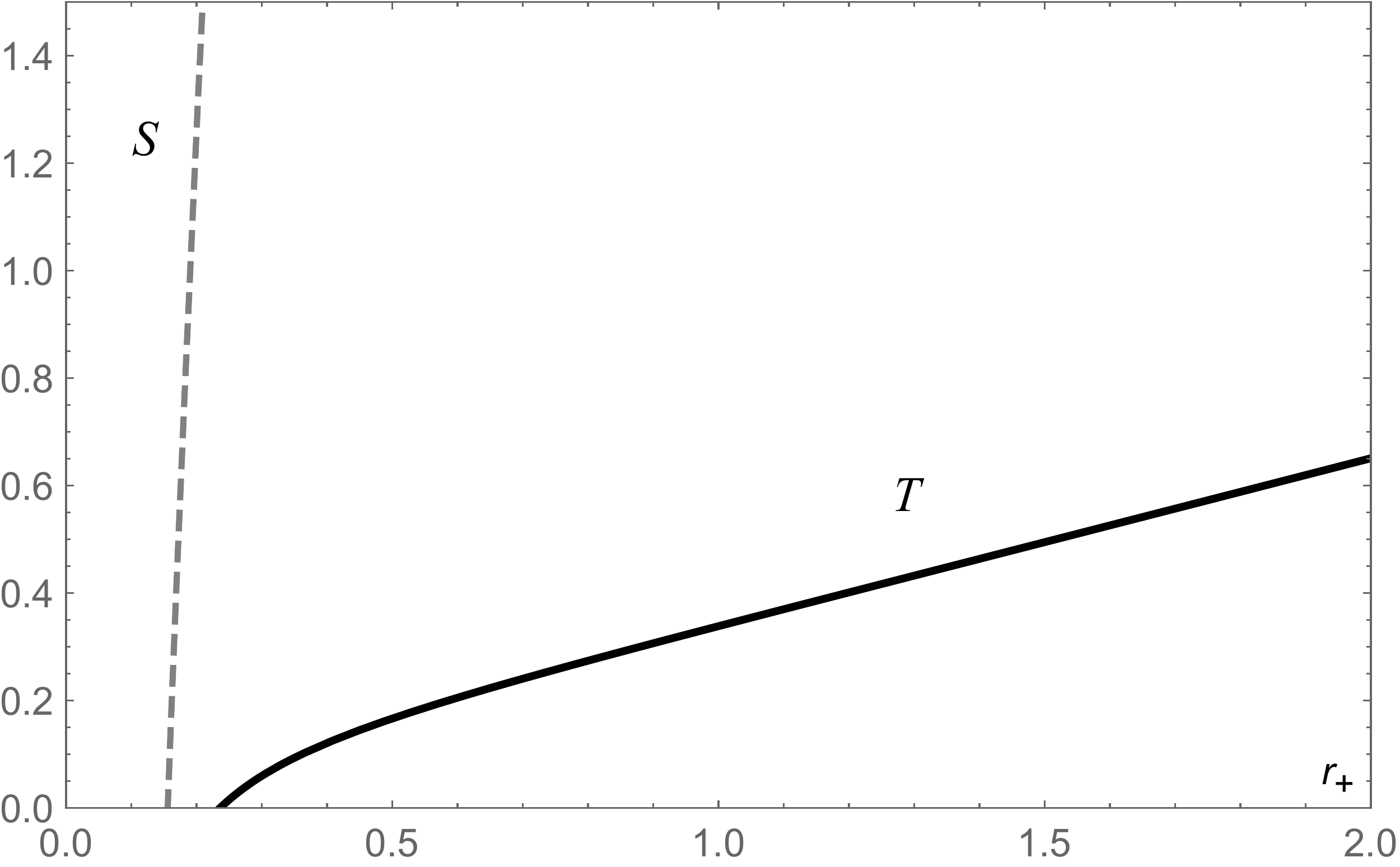}
\caption{These graphs show the temperature $T$ and the entropy $S$ as functions of the horizon radius $r_+$. Two qualitatively different behaviours are shown, and these correspond to different choice of the dimensionful coupling constants. In particular, for these plots we have chosen $a_2>0$, $a_3<0$ and $q<0$.}
\end{figure}
\end{center}

\section{Adding $R^4$ terms}

It was observed in \cite{MD, DV} that in 5 dimensions one can construct a
higher-derivative combination which is quartic on the Riemann tensor, and that, as quasi-topological gravity, leads to second order
field equations when evaluated on a spherically symmetric ansatz. As pointed out in
\cite{MD}, even thought the geometric meaning of such term remains unclear, it
serves as a toy model to study quartic corrections in 5 dimensions in a
controlled setup. This theory was dubbed quartic quasi-topological gravity and,
as its cubic counterpart, leads to the Einstein propagator when perturbed
around a (A)dS.

In this section we show that this theory with $R^4$ terms also admits hairy black hole
solutions. Due to the fact that terms of higher order would generically lead to quintic Wheeler polynomials, the solution
we will provide here can be regarded as the most general of its class that allows to be solved algebraically. This amounts to augment the Lagrangian (\ref{La4}) with a term 
\[
\Delta I^{(4)} = \frac{a_{4}}{73\times2^{5}\times3^{2}}\int d^5x \sqrt{-g} \ \mathcal{L}_{4}%
\]
with the quartic Lagrangian density given by
\begin{align} \label{PoneteUnaCorbata}
\mathcal{L}_{4}  &  =7080R^{pqbs}R_{p\ b}^{\ a\ u}R_{a\ u}^{\ v\ w}%
R_{qvsw}-234R^{pqbs}R_{pq}^{\ \ au}R_{au}^{\ \ vw}R_{bsvw}-1237\left(
R^{pqbs}R_{pqbs}\right)  ^{2}\\ \nonumber
&  +1216R^{pq}R^{bsau}R_{bs\ p}^{\ \ v}R_{auvq}-6912R^{pq}R^{bs}%
R_{\ p\ q}^{a\ u}R_{abus}-7152R^{pq}R^{bs}R_{\ \ pb}^{au}R_{auqs}\\ \nonumber
&  +308R^{pq}R_{pq}R^{bsau}R_{bsau}+298R^{2}R^{pqbs}R_{pqbs}+12864R^{pq}%
R^{bs}R_{b}^{\ a}R_{psqa}-115R^{4}\\ \nonumber
&  -912RR^{pq}R^{bs}R_{pbqs}+4112R^{pq}R_{p}^{\ b}R_{q}^{\ s}R_{bs}%
-4256RR^{pq}R_{p}^{\ b}R_{qb}+1156R^{2}R^{pq}R_{pq}\ . 
\end{align}

Notably, this theory also admits hairy black holes of the type described above. The configuration for the scalar and gauge field remain (\ref{La15}) and (\ref{La16}) respectively. The metric function $f\left(  r\right)  =g\left(  r\right)  +k$ is now given by the roots of the polynomial
\begin{equation}
a_{4}g^{4}+2a_{3}r^{2}g^{3}+12a_2 r^{4}g^{2}-6r^{6}g-\Lambda r^{8}%
-6(2a_2 |k|+\mu)r^{4}-6qr^{3}+6r^{2}e^{2}=0\ . \label{Solution4}
\end{equation}
from what we can see that the curvature of the constant curvature solution
with $f\left(  r\right)  =-\frac{\lambda}{6}r^{2}+k$, fulfills%
\begin{equation}
a_{4}\lambda^{4}-12a_{3}\lambda^{3}+432a_2\lambda^{2}+1296\left(
\lambda-\Lambda\right)  =0\ ,
\end{equation}
provided%
\begin{equation}
\mu=-2|k|a_2\ .
\end{equation}

The perturbations around the AdS$_5$ vacuum not to contain ghosts demands the constraint
\begin{equation}
1-\frac{2}{3} a_2 |\lambda | - \frac{1}{36} a_3 |\lambda |^2  - \frac{1}{324} a_4 |\lambda |^4 > 0 \ ,
\end{equation}
cf. (\ref{noghosts}). 


The mass of the general solution (\ref{Solution4}) as a function of the horizon radius can be computed in a similar way as before. It is given by
\begin{equation}
M  =\frac{\text{Vol}}{2r_{+}^{4}}\left(  -\Lambda r_{+}^{8}+6kr_{+}^{6}-6qr_{+}^{3}-2a_{3}k%
r_{+}^{2}+6e^{2}r_{+}^{2}+a_{4}|k|\right)  -M_{0}\ ,
\end{equation}
where we have made explicit the background gravitational energy $M_0$. The temperature and the entropy, on the other hand, take the form 
\begin{equation}
T=\frac{1}{4\pi r_{+}}\frac{-2\Lambda r_{+}^{8}+6kr_{+}^{6}+3qr_{+}^{3}+2(a_{3}k-3e^{2}%
)r_{+}^{2}-2a_{4}|k|}{(3r_{+}^{6}+12a_2 kr_{+}^{4}-3a_{3}|k|r_{+}^{2}+2a_{4}k)}\ .
\end{equation}
and
\begin{equation}
\frac{S}{\text{Vol}}=4\pi r_{+}^{3}+48\pi r_{+}ka_2
-10\pi k q +\frac{12\pi a_{3}|k|}{r_{+}}
-\frac{8\pi
a_{4}k}{3r_{+}^{3}}\ .
\end{equation}
respectively, which can also be seen to fulfil the first principle (\ref{Elprimero}). 

\section{Quasi-topological matter}

With the purpose of studying to what extent we can generalize the black hole solutions described above, in this section we will consider a more general type of coupling with the scalar field $\chi $. This will include quadratic, cubic, and quartic couplings with the Riemann tensor. However, this will differ at cubic and quartic order from other couplings considered in the literature previously \cite{OR, OR2}. In fact, the theory we will consider here is the natural conformal generalization of quasi-topological gravity: Consider the theory defined by coupling the Einstein-Hilbert action with cosmological term to the matter action
\begin{equation}
I_{\text{matter}}\left[  g_{ab},\chi \right] = ç \sum_{n=0}^{4}  
\ b_n \int d^5x \sqrt{-g}  \ t_{a_1 b_1 ... a_n b_n }^{c_1 d_1 ... c_n d_n } \ \hat{R}_{c_1 d_1}^{\ \ \ \ a_1 b_1} ...\ \hat{R}_{c_n d_n}^{\ \ \ \ a_n b_n} \ , \label{Mingus}
\end{equation}
where $\hat{R}_{a\ cd}^{\ b}$ is defined in (\ref{Stensor}), and depends on the scalar $\chi $. Tensor $t_{a_1 b_1 ... a_n b_n }^{c_1 d_1 ... c_n d_n }$ for $n\leq 2$ is taken to be (\ref{JKL}), while for $n=3$ and $n=4$ it coincides with the ones that yield the cubic term in (\ref{L3OR}) and the quartic term in (\ref{PoneteUnaCorbata}), respectively. It is in this sense that action (\ref{Mingus}) can be regarded as the natural generalization of quasi-topological gravity. Indeed, (\ref{Mingus}) is to quasi-topological gravity what the theory introduced in \cite{PutoElQueLee} is to Lovelock gravity.

Remarkably, GR coupled to matter theory (\ref{Mingus}) also admits black hole solutions whose metrics we can write down explicitly. If for simplicity we consider the case in which the coupling constants $b_n$ obey the following relations
\begin{equation}
3b_1n^{30}+12b_2kn^{20}-3b_3|k|n^{10}+2b_4k= 0 \ , \\ \label{CouplingMagico}
\end{equation}
\begin{equation}
-5b_0n^{40}-12b_1kn^{30}+24b_2|k|n^{20}-4b_3kn^{10}-2b_4|k|= 0 \ , \label{CouplingMagico2}
\end{equation}
the theory admits solutions of the form (\ref{anz}) with $N=1$ and with the function $f(r)$ given by
\begin{equation}
f(r)=-\frac{\Lambda r^2}{6}+1-\frac{m}{r^2}-\frac{q}{r^3} \ , \label{Poliamorrr}
\end{equation}
where $\mu $ is an integration constant and $q$ is given by 
\begin{equation}
q=\frac{1}{3n^{15}}\left(b_0 n^{40}+6b_1kn^{30}+2b_3kn^{10}-2b_4|k|\right) \ , \label{laqk} 
\end{equation}
with the scalar field $\chi $ being (\ref{La15}). It is worth pointing out that the conditions (\ref{CouplingMagico})-(\ref{CouplingMagico2}) happen to be less restrictive than conditions such as (\ref{La19}), cf. \cite{GLOR}. 

As we said, since the matter action (\ref{Mingus}) differs from other higher-curvature couplings considered in the literature \cite{OR, OR2, Tanhayi:2012nn}, the black hole solution we find from (\ref{Poliamorrr})-(\ref{laqk}) also differs from the solutions studied in \cite{GLOR, Nosotros}.

\section{Quasi-topological gravity with a unique vacuum}
As in Lovelock theory, for a given value of the couplings, the theory admits a unique maximally symmetric solutions of constant curvature $-1/l^2$. Such theory admits black hole solutions with simple black hole metrics \cite{grandecamarada}-\cite{grandecamarada2}, that can be written in a closed form. As shown in \cite{OR}, cubic quasi-topological gravity also admits such black holes. Below we show that even in the presence of conformal matter and a Maxwell term, cubic and quartic quasi-topological gravities admit a simple black hole metric when the couplings are such that there is a unique maximally symmetric solutions. Requiring the theory to have a unique vacuum with curvature $-1/l^2$ implies that the gravitational action can be written as
\begin{equation}
I_{\text{gravity}}\left[  g_{\mu\nu} \right] =\kappa\int d^{5}x \sqrt{-g} \left( a_{0}+a_{1}R+a_{2}\mathcal{L}_{2}  +a_{3}\mathcal{L}_{3} + a_{4}\mathcal{L}_{4}\right)  \, ,  \label{lcg}
\end{equation}
provided
\begin{equation}\label{cubiccoeffs}
a_{0}=\frac{12}{l^6}\,,\quad  a_{1} = \frac{3}{l^4}\,,\quad a_{2} = \frac{3}{2l^2}\,,\quad a_{3} = -3\,, \quad a_{4}=0\,.
\end{equation}
when the quartic term is not present, or
\begin{equation}\label{cuarticcoeffs}
a_{0}=\frac{12}{l^8}\,,\quad a_{1} = \frac{4}{l^6}\,,\quad a_{2}=\frac{3}{l^4}\,,\quad a_{3} =-\frac{12}{l^2}\,,\quad a_{4} = 6\,\,.
\end{equation}
when the quartic quasi-topological term is non-zero. The global constant $\kappa$ has mass dimension $-(2p-5)$ where $p$ is the maximum power of the curvature that appears in the Lagrangian. In these cases, the charged black hole solution with conformal hair is defined by the following lapse function
\begin{equation}\label{ftunned}
f(r)=k+\frac{r^2}{l^2}-(\pm)^{p+1}\kappa^{-1/p}\left(\mu r^{2p-4}+qr^{2p-5}-e^2r^{2p-6}\right)^{1/p}
\end{equation}
where $k=\pm 1,0$ is the curvature of the horizon, as above. The matter fields are given by (\ref{La15}) and (\ref{La16}) with the relations (\ref{La19}), (\ref{elq}) and (\ref{eln}). Here $\mu$ is an integration constant, which is related to the mass. Notice that the presence of the $p^{\text{th}}$ root in (\ref{ftunned}) demands the horizon $r_+$ to be larger than the critical value $r_{\text{c}} = (|q|+\sqrt{q^2+4\mu e^2} )/(2\mu )$ at which the radicand vanishes.

\section{Concluding remarks}

In this paper we have studied charged AdS black holes in quasi-topological higher-derivative gravity. Our solution generalizes those found and studied previously in references \cite{OR, MR, DV}, and show that black hole solutions in 5-dimensional quasi-topological gravity can be equipped with a scalar hair which backreacts on the metric. The metric of these hairy black holes can be written analytically in the case when both $R^3$ and $R^4$ terms are present. The scalar field configuration does not endow the black hole with new quantum numbers, but rather exhibits a fixed intensity, and therefore describes a secondary hair. The scalar field remains finite everywhere outside and on the horizon, and it behaves at large distance in a way that respect the asymptotic AdS$_5$ boundary conditions relevant for AdS/CFT.

The solutions we described here exist in the region of the parameters space where consistency conditions are fulfilled. In particular, the linearized theory around AdS$_5$ exhibits massless spin-2 excitations with the adequate sign of the kinetic term. Other consistency conditions, such as the bounds on the central charges ratio $a/c$ in the dual boundary theory can be satisfied. 

The thermodynamical properties of the hairy black hole solutions can be studied analytically provided we have derived explicit expressions for the conserved charges, the Hawking temperature, and the black hole entropy. While in the IR limit those formulas reproduce the thermodynamics of GR black holes in AdS$_5$, the higher-curvature terms induce corrections in the UV. In particular, new qualitative features appear, such as remnants and extremal configurations. In particular, this implies that black holes with arbitrarily low temperature exist in AdS$_5$. The theory also allows for the existence of phase transitions \cite{GGO, Mann2, Mann3} involving hairy black holes. As mentioned in the introduction, this provides an interesting setup to study AdS/CFT. In particular, it is interesting to investigate what is the interpretation of the scalar hair from the holographic point of view.


\[ \]
\subsection*{Acknowledgments}

The authors thank Sourya Ray and Andr\'es Goya for useful discussions and previous collaboration on this subject.  We also thank G.A. Silva for references. G.G. and J.O. are grateful to Alberto G\"uijosa and Mariano Chernicoff for their hospitality during their visit to UNAM, where this work was finished. This work has been supported by FONDECYT
Grants 1141073 and 1150246 and by Newton-Picarte Grant DPI20140053. MC is partially supported by Mexico's National Council of Science
and Technology (CONACyT) grant 238734, DGAPA-UNAM grant IN113115 and IN107115. G.G. is partially supported by CONICET through the grant PIP 0595/13, and by NSF/CONICET bilateral cooperation program.

\[ \]

\end{document}